\documentclass{article}
\usepackage[utf8]{inputenc}
\usepackage{booktabs}
\usepackage{graphicx}
\usepackage{subcaption}
\usepackage{xcolor}
\usepackage{color}
\usepackage{soul}
\usepackage{hyperref}
\usepackage[round]{natbib}
\setcitestyle{numbers}
\usepackage{multirow}
\usepackage[T1]{fontenc}

\begin{document}

\title{Cross-Platform Social Dynamics: An Analysis of ChatGPT and COVID-19 Vaccine Conversations}
\author{Shayan Alipour$^{a,*}$ $\,\,$ Alessandro Galeazzi$^b$ $\,\,$ Emanuele Sangiorgio$^{c}$ \\ Michele Avalle$^{a}$ $\,\,$ Ljubisa Bojic$^{d,e}$ $\,\,$ Matteo Cinelli$^{a}$ \\ Walter Quattrociocchi$^{a}$}

\vspace{-6ex}
\date{}

\maketitle

\begin{center}
{$^a$\textit{\small Sapienza University of Rome --- Department of Computer Science}\\
$^b$\textit{\small Ca'Foscari University of Venice, DAIS, Venice, Italy} \\
$^c$ \textit{\small Sapienza University of Rome --- Department of Social and Economic Sciences} \\
$^d$\textit{\small The Institute for Artificial Intelligence Research and Development of Serbia} \\
$^e$\textit{\small University of Belgrade, Institute for Philosophy and Social Theory}\\ 
$^*$ \textit{\small Correspondence should be addressed to: shayan.alipour@uniroma1.it}
}
\end{center}

\begin{abstract}
The role of social media in information dissemination and agenda-setting has significantly expanded in recent years. By offering real-time interactions, online platforms have become invaluable tools for studying societal responses to significant events as they unfold.
However, online reactions to external developments are influenced by various factors, including the nature of the event and the online environment. This study examines the dynamics of public discourse on digital platforms to shed light on this issue. We analyzed over 12 million posts and news articles related to two significant events: the release of ChatGPT in 2022 and the global discussions about COVID-19 vaccines in 2021. Data was collected from multiple platforms, including Twitter, Facebook, Instagram, Reddit, YouTube, and GDELT.
We employed topic modeling techniques to uncover the distinct thematic emphases on each platform, which reflect their specific features and target audiences. Additionally, sentiment analysis revealed various public perceptions regarding the topics studied. Lastly, we compared the evolution of engagement across platforms, unveiling unique patterns for the same topic.
Notably, discussions about COVID-19 vaccines spread more rapidly due to the immediacy of the subject, while discussions about ChatGPT, despite its technological importance, propagated more gradually.
\end{abstract}

\section*{Introduction}

Social media have markedly reshaped global information access, sharing, and consumption, thereby redefining the dynamics of information dissemination and, consequently, agenda-setting dynamics~\cite{schmidt2017anatomy, quattrociocchi2014opinion, del2016spreading}. The spread and consumption of information on online social media may be influenced by several factors such as biases~\cite{messing2014selective,cinelli2020selective}, platform designs, and algorithms~\cite{etta2023characterizing,cinelli2021echo}. Typically, online users are inclined towards information that resonates with their viewpoints~\cite{bessi2015science}, often dismissing opposing data~\cite{zollo2017debunking}, leading to the formation of like-minded user groups supporting a common narrative~\cite{del2016spreading}. The dynamics of these interactions may vary across social media platforms due to differences in business models and content selection algorithms~\cite{valensise2023drivers}. Online discourse frequently centers around controversial or timely topics such as political elections~\cite{kubin2021role,bail2018exposure}, natural events~\cite{falkenberg2022growing}, or significant global occurrences~\cite{briand2021infodemics}.

Recent advancements in Large Language Models (LLMs) have attracted significant attention due to their potential impact on various sectors \cite{radford2019language}. These models, trained on extensive datasets, can process and generate text similar to human communication, exhibiting quick and effective adaptability to new tasks \cite{llmIntro}. A notable instance is ChatGPT, launched by OpenAI on November 30, 2022\footnote{\url{https://openai.com/blog/chatgpt}}, which has catalyzed discussions regarding its capabilities and associated risks, including misinformation, ethical considerations, and broader AI implications \cite{ray2023chatgpt}.
LLMs like ChatGPT have displayed remarkable competence in diverse tasks, ranging from creative writing to complex problem-solving \cite{aimodelreviews, gilardi2023chatgpt}, and their usage has proliferated across various professional domains and among the general public. In particular, the accessibility of ChatGPT to the general public triggered a substantial volume of posts across multiple social media platforms shortly after its release \cite{explodingtopics}. While many discussions were positive, growing concerns regarding the potential risks such as misinformation dissemination, cybersecurity threats, and adverse impacts on the labor market also fueled the discourse \cite{hallucination, chatgpt_security_risk, threatGPT}.
These concerns have also given rise to alternative discussions emphasizing the limitations of LLMs in precise planning and problem-solving \cite{dziri2023faith, valmeekam2022large, mahowald2023dissociating}.

The COVID-19 pandemic emphasized the challenges posed by information saturation and highlighted the central role of social media in its dissemination \cite{zarocostas2020fight,cinelli2020covid, briand2021infodemics}. Analyzing the spreading patterns of innovations like ChatGPT across digital platforms, and comparing these dynamics with other significant events, is crucial for evaluating the social media platforms' impact in our society.

In this study, we investigate the trajectory of ChatGPT discourse across online platforms, using data from five major social media platforms — Facebook, Twitter, Instagram, Reddit, and YouTube — alongside global news coverage captured by the GDELT dataset. We capture user engagements, tracing the rise in interest and participation across diverse platforms while characterizing the debate surrounding LLMs by identifying dominant themes and sentiments. Additionally, we model the growth trajectory of user engagement within the LLM discourse and compare it with the user growth pattern related to COVID-19 vaccination discussions, a well-documented controversy \cite{cinelli2020covid}. This comparison aims to elucidate the differences in information dissemination dynamics across global topics.

Although some recent studies have looked into online discussions about ChatGPT \cite{whatChatGPTdo, leiter2023chatgpt, early2022adapt, koonchanok2023tracking}, they did not provide a comparison across different social media or considered global news coverage. The nature of debates can change based on the platform they occur on \cite{cinelli2021echo,cinelli2020covid}, so analyzing discussions in different online settings is crucial to gaining a thorough understanding. Our analysis fills this gap by examining discussions on multiple social media platforms and news outlets.

In this study, we identified a concise set of relevant topics from comments about LLMs, like risks, health, education, finance, and technical discussions. We found that users on different platforms focus markedly on different topics, reflecting the distinctive nature of each platform. By performing sentiment analysis, we were also able to identify specific themes that most represent concern and excitement toward the recent deployment of AI. When we modeled the user growth pattern on each platform, we found that users engaged faster with discussion about ChatGPT on Twitter, YouTube, and Reddit compared to Facebook and Instagram. We also noticed that COVID-19 vaccine debates spread faster than those about ChatGPT on all platforms. In both cases, discussions on social media spread faster than in news articles.

Our research underscores the importance of understanding online discussions within their unique contexts. It highlights the factors affecting how information spreads across various platforms and topics. The findings from our study have implications for how we perceive the spread of information online, especially during critical global events.
Our findings are also in line with key communication and media theories such as selective exposure theory~\cite{sears1967selective,stroud2008media}, the agenda-setting function of media~\cite{mccombs1972agenda,scheufele2007framing} and the role of framing in decision making~\cite{entman1993framing, chong2007framing}. Evidence that people consume information that aligns with their existing beliefs and attitudes, posited by selective exposure theory, emerges from our results as users on different platforms gravitated towards discussing aspects of ChatGPT that interest them or align with their perspective. Moreover, the agenda-setting theory suggests that the media significantly influences what issues are important to the public based on the coverage they receive. The heightened discussion around ChatGPT across all platforms following its release is a classic example of this theory in play. Lastly, the importance of framing is also evident in our research. How topics related to LLMs and COVID-19 vaccines are presented on different platforms can significantly affect perceptions and resultant user engagements. Analyzing these frames can provide insights into how these topics can be most effectively communicated.
By recognizing the diverse focus and engagement patterns on different platforms, stakeholders, including policymakers, educators, and the tech industry, can better anticipate and respond to public reactions and concerns in a digitally connected world.

\section*{Results}

\subsection*{ChatGPT Discussion in Online Platforms}

\begin{figure}[ht!]
    \centering
    \includegraphics[width=0.99\textwidth]{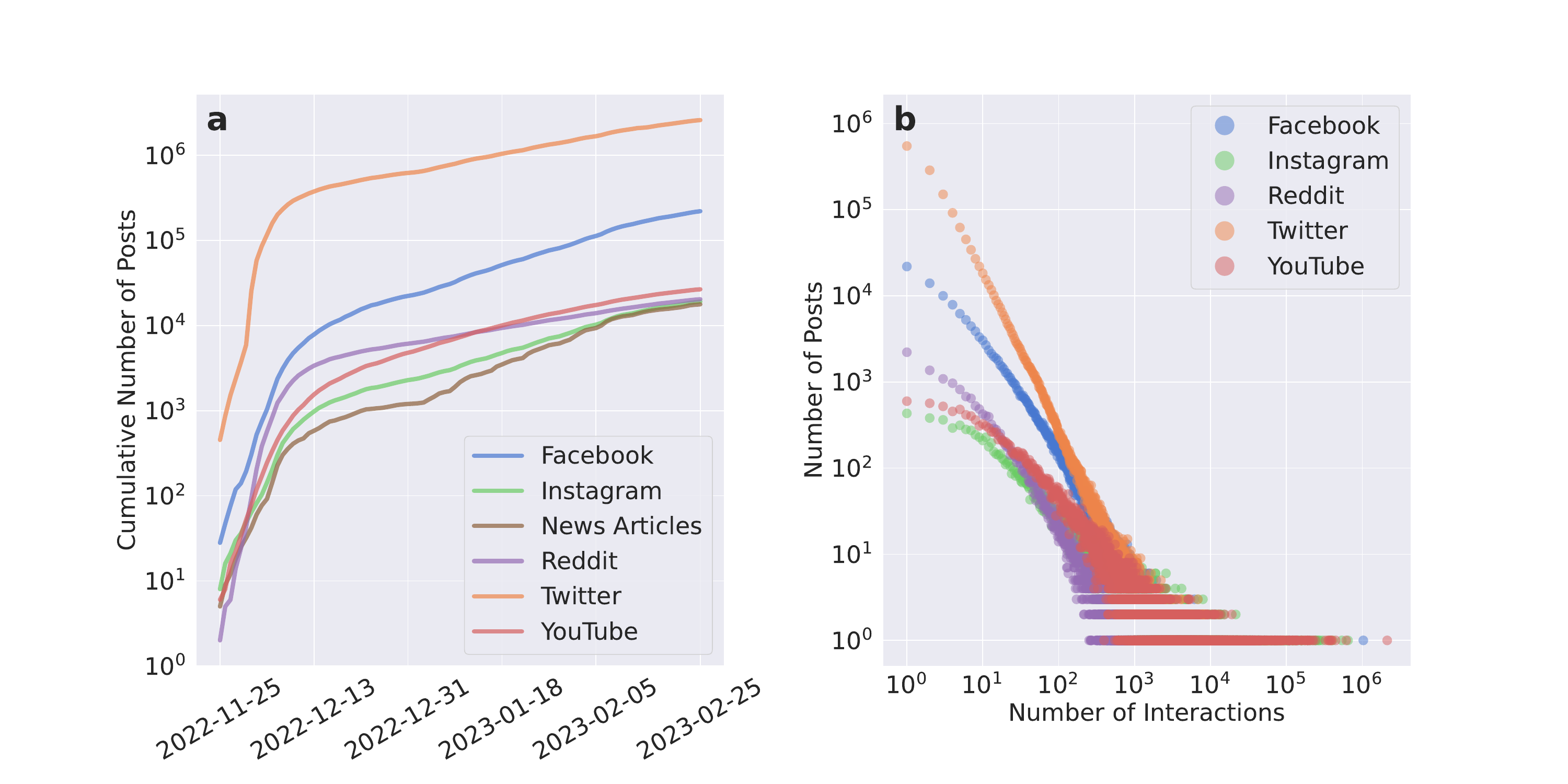}
    \caption{Cumulative number of unique posts about ChatGPT discussion across various platforms (a) and distribution of interaction volume versus the number of posts on different platforms (b). The nature of interactions varies among platforms; For instance, on Twitter, interactions are the sum of likes, quotes, retweets and replies, while on Instagram and YouTube, interactions are the sum of likes and comments.}
    \label{fig:general_info}
\end{figure}

We begin our analysis by outlining the discussion around ChatGPT across different platforms. Our dataset includes about 3 million news articles and posts from November 25, 2022, to February 25, 2023 (see Methods for more details). Figure \ref{fig:general_info}.a displays the cumulative count of content related to ChatGPT over three months, highlighting a sharp increase in early December followed by steady growth. Twitter and Facebook are identified as the most active platforms, while the discussion has a lower volume of news articles coverage. Figure \ref{fig:general_info}.b shows the distribution of interactions by platform, where interactions consist of likes, comments, shares, and platform-specific metrics. Despite differences in platforms, all interaction distributions exhibit a long tail pattern consistent with previous studies~\cite{cinelli2020covid}, indicating that a small number of posts receive the most interactions while the majority receive minimal consideration.
This pattern of engagement confirms the skewed nature of online discussions, where only a few posts dominate the conversation and receive disproportionate attention across different social media platforms.

\subsection*{Platform-Specific Dynamics}
\begin{figure}[ht!]
    \centering
    \includegraphics[width=0.99\textwidth]{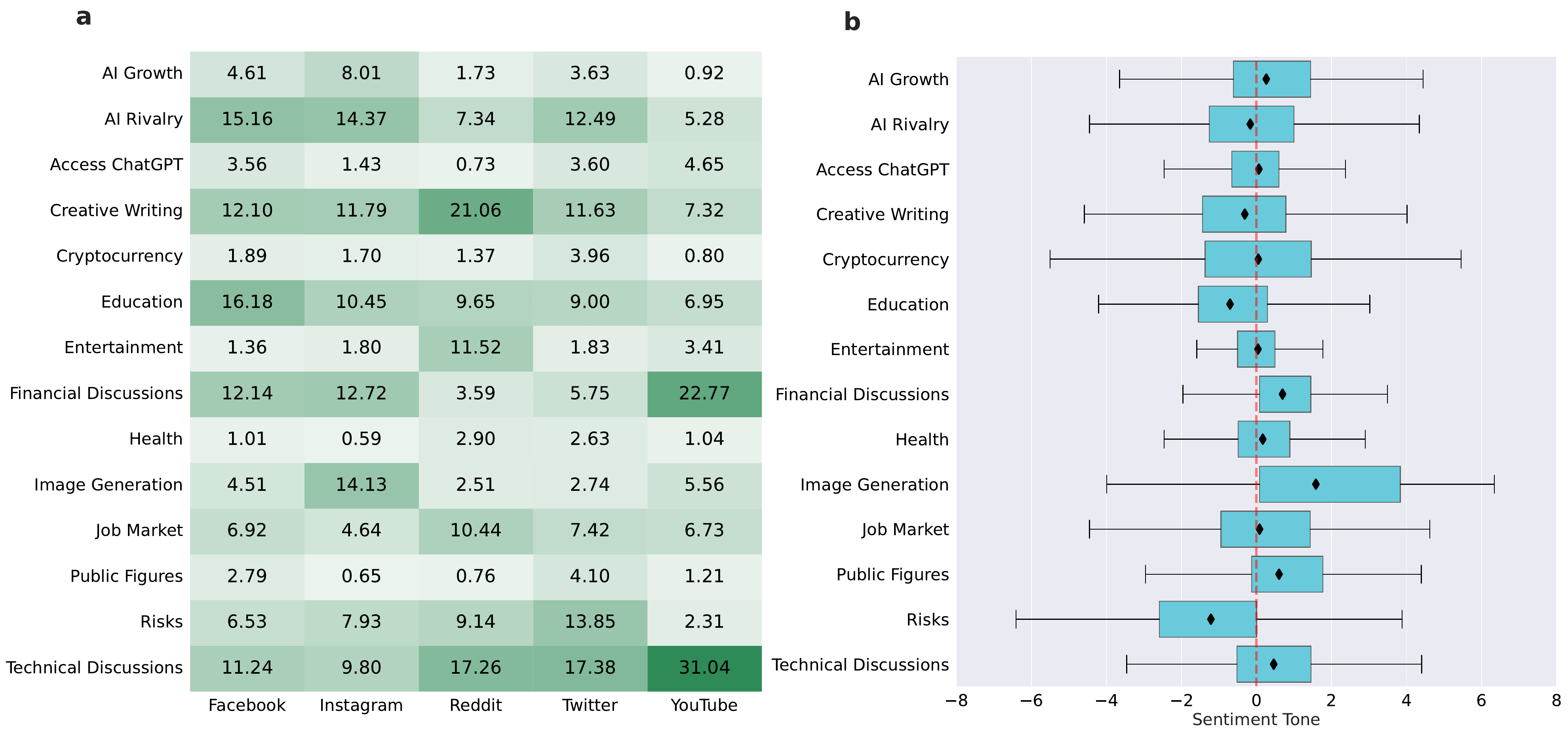}
    \caption{(a) Proportion of comments for each topic by platform. 
    The cell color intensity corresponds to the proportion of comments discussing a given topic; a higher percentage results in a darker hue.
    (b)  Box plots distributions of the sentiment tone across topics. On the x-axis, sentiment tones are represented as values. A negative value indicates a negative sentiment, while a positive value suggests the opposite. The further away from zero the value is, the stronger the sentiment. The vertical red dashed line at the 0 mark, differentiates positive tones from negative tones. Black diamonds inside the boxes indicate the average sentiment tone for each topic.}
    \label{fig:topics}
\end{figure}

In this section, we apply topic modeling techniques to identify the main themes in ChatGPT discussion across platforms (see Methods for more details).
Figure~\ref{fig:topics}.a reports the percentage of comments discussing different topics on each platform, revealing that the discourse surrounding ChatGPT follows distinct patterns that vary from platform to platform, potentially reflecting differences in their user bases. 
For instance, Instagram stood out for its significant attention to image generation, which discusses tools like Midjourney, Stable Diffusion, and DALLE-2 for crafting visuals. The role of AI in education grabbed the most attention among users on Facebook, touching upon its implications for plagiarism, how schools might incorporate LLMs, and the evaluation of AI in academic settings.
Facebook and Instagram saw a large debate about how ChatGPT can be used in the context of finance (i.e. ``Financial Discussions'') but, in this regard, they were outpaced by YouTube, where investments and (personal) finance topics usually get a large share of interest \cite{burgess2018youtube}.  
``Creative Writing'' was a major topic for Reddit users, where users mainly discussed ChatGPT's ability in various writing tasks like poetry, songs, screenplays, and emails.
``Technical Discussions'', which covers AI tutorials, LLM training, and integration, seemed to resonate more with users on YouTube, Twitter, and Reddit, suggesting a user base eager to discuss the working mechanism of these language models.
Despite the observed difference, the implications of AI on the job market emerged as a consistent theme across platforms. In the topic of ``Job Market'', comments covered the potential increased productivity while also addressing concerns about human job replacement.
Users on Twitter relatively discussed more the topic related to the potential risks associated with LLMs. In summary, the topic of ``Risks'' is about posts discussing issues from data misuse, jailbreaking, potential biases, and societal impacts of AI.
Finally, we note how the topic ``Health'' got less interest in the early weeks since the launch of ChatGPT. This lack of interest is surprising considering present public health community concerns about its role in substituting healthcare experts~\cite{javaid2023chatgpt} and other global phenomena such as Infodemic~\cite{infodemic2023chatgpt}.

Further, we analyzed the sentiment tone distribution around topics by leveraging data from Global Database of Events, Language, and Tone (GDELT), a comprehensive resource that systematically gathers news content from various news outlets \cite{leetaru2013gdelt, leetaru2015mining}. We tied the sentiment tone of news articles to the comments that mentioned them, using these articles as a proxy for the sentiment tone of the comments. Figure \ref{fig:topics}.b shows the sentiment distribution for each topic. The breakdown of distribution statistics is available in table \ref{tab:tone_stats}. We obtained further evidence about both public concern and enthusiasm regarding the use of ChatGPT, showing that AI's effects on education and writing are perceived as negative. On the other hand, ChatGPT was mostly discussed in a positive way in image generation, financial, and technical conversations. Nonetheless, the widespread sentiment distribution is common to all the extracted topics and underlines the potential controversy it generated in the public debate.

\subsection*{Modeling User Engagement Growth across Platforms}

After identifying differences and similarities in ChatGPT discussions across various platforms and noting their potential to spark debate, we aim to track how new users post ChatGPT-related content. By examining how the number of users grows over time, we can quantify the differences between platforms in the evolution of the discourse.

We processed the data to consider the cumulative number of unique users up to each day, counting each user that participated in the debate once, marked on their first appearance day. We modeled users' growth using logistic function (for more details about the model refer to the Methods section). The model's main coefficients, $\alpha$ and $\beta$, represent the growth rate and the time half of the unique users were engaged, respectively. A higher growth rate indicates a more steep increase in users' volume and is quantified by a higher $\alpha$ value, while a lower $\beta$ value implies that it takes a shorter time to engage 50\% of the final user base.
We also measured the growth speed through the Speed Index (\textit{SI}) at which the model reaches its plateau and can be interpreted as how fast the discussion saturates among users. The speed index provides a comparative measure of user engagement dynamics across platforms because of its normalized value. Figure~\ref{fig:fits} depicts a series of plots showing the cumulative sum of unique users engaged in ChatGPT-related topics for each platform, while for news articles, it shows the cumulative sum of published articles present in the GDELT dataset. In each plot, we report the curve obtained by fitting the logistic function Eq.~\ref{eq:sigmoid} to model the users' growth, while the parameters of the fits ($\alpha$, $\beta$, and \textit{SI}) are detailed in Table \ref{tab:fit_vars}. 
Twitter, YouTube, and Reddit exhibit similar growth rates ($\alpha$ values) and times to reach half of the unique users ($\beta$ values). They also have higher Speed Index values with respect to other platforms, indicating faster growth in unique users compared to Instagram and Facebook.
Conversely, Instagram and Facebook demonstrate steeper user growth (higher $\alpha$ values) but require more time to engage half of the unique users (higher $\beta$ values). 
Remarkably, Facebook has the second largest user volume, but a SI lower than all other platforms except Instagram, suggesting that a delayed interest in the user base is independent of the size.
Regarding news articles, it seems that social media users are leading the way in content generation during this period, a finding that aligns with previous studies underscoring the leading role of social media in shaping the news landscape\cite{socialMediaBetterNews, newsdiffusion}. 

We observed platforms may exhibit different user engagement patterns as the discussion on a topic evolves. Understanding these dynamics is crucial for planning the dissemination of information and managing online discussions about major events.

\subsection*{Comparing ChatGPT Discourse with COVID-19 Vaccine Discussions}
\begin{figure}[ht!]
    \centering
    \includegraphics[width=0.999\textwidth]{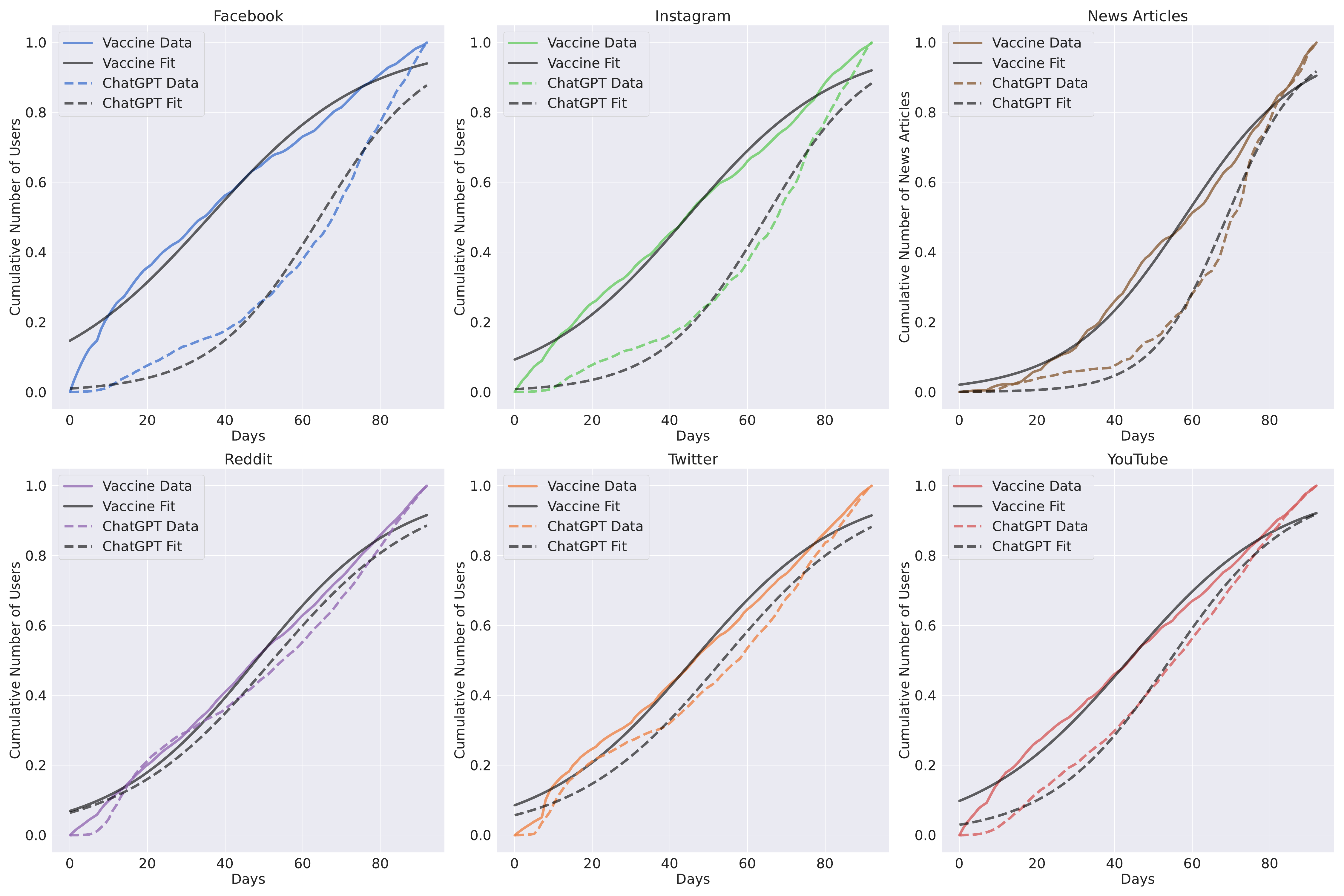}
    \caption{Cumulative number of unique users with logistic fits by platform. Each plot shows the cumulative count of unique users engaged in ChatGPT and COVID-19 vaccination-related topics over time. The fitted curve corresponds to a logistic function used to model the diffusion of unique users.}
    \label{fig:fits}
\end{figure}

To further clarify the impact of ChatGPT in the public discourse on various platforms we compare its growth pattern with another topic that got significant attention, namely discussions surrounding COVID-19 vaccines.
The discourse on vaccines, inflated by urgent health concerns during the pandemic and spread quickly across various media platforms\cite{valensise2021lack}. Conversely, ChatGPT represents a different kind of subject, being a notable technological advancement of potentially comparable resonance~\cite{chatgptPandemic,infodemic2023chatgpt}. This analysis compares the rates at which different content spreads across diverse social media platforms. Additionally, we include news articles in the analysis to grasp ChatGPT's impact on global media coverage.
To carry out this comparison, a dataset on COVID-19 vaccine discussions was curated across the same platforms and within a comparable time range as our ChatGPT dataset. This specific timeframe was selected due to the substantial debates and conversations surrounding COVID-19 vaccines, making it a suitable benchmark for comparison with the ChatGPT discourse. Figure~\ref{fig:fits} displays a series of plots showcasing the cumulative count of unique users engaged in vaccine-related discussions across each platform, while for news articles, it illustrates the cumulative total of published pieces. Accompanying each plot is a curve representing the fitted logistic function, providing a model for the diffusion process underlying the growth of the number of unique users in the vaccine debate. Table~\ref{tab:fit_vars} reports the parameters for fitting these logistic curves. The fitted results of ChatGPT-related data are also included for a visual comparison.
The normalized curves account for variations in user base sizes and allow for a direct comparison across platforms.
In light of the comparative analysis, several key takeaways emerge regarding the evolution of ChatGPT and COVID-19 vaccine debates across different platforms. In all platforms we observed that COVID-19 vaccine debate exhibits a considerably higher Speed Index than the ChatGPT discourse. Notably, more than the others, Twitter and Reddit exhibit a higher similarity in the engagement patterns between the two topics. 
This disparity underscores the faster and wider spread of information about COVID-19 vaccines with respect to ChatGPT discourse.
This aligns with previous research, which found that online users are prone to engage with highly controversial topics such as climate change or health-related subjects~\cite{falkenberg2022growing,cinelli2021echo}.
Moreover, the analysis underlines the dependency of information spreading patterns on the type of environment and audience, with platforms' user bases showing different interest levels in various topics.

\begin{table}[ht!]
\centering
\caption{Logistic function fitting parameters for various datasets. The table reports the $\alpha$ (growth rate), $\beta$ (point of half-saturation), and \textit{SI} (Speed Index) values, which are derived from fitting the data for each platform into a logistic function. \textit{SI} measures the normalized area under the curve. Root mean square error (RMSE) quantifies the model accuracy in fitting data. The smaller the value, the closer the predictions are to the observations.}
\label{tab:fit_vars}
\begin{tabular}{c c c c c c}
\toprule
\textbf{Topic} & \textbf{Platform} & \textbf{$\alpha$} & \textbf{$\beta$} & \textbf{\textit{SI}} & RMSE \\
\midrule
\multirow{6}{*}{ChatGPT} & Facebook & 0.071 & 64.396 & 0.318 & 0.042 \\
 & Instagram & 0.074 & 64.716 & 0.314 & 0.041 \\
 & News & 0.104 & 68.858 & 0.260 & 0.034 \\
 & Reddit & 0.051 & 52.125 & 0.445 & 0.049 \\
 & Twitter & 0.052 & 53.524 & 0.432 & 0.048 \\
 & YouTube & 0.064 & 54.198 & 0.420 & 0.030 \\
\midrule
\multirow{5}{*}{Vaccine} & Facebook & 0.049 & 35.855 & 0.589 & 0.038 \\
 & Instagram & 0.051 & 44.350 & 0.515 & 0.034 \\
 & News & 0.066 & 57.941 & 0.383 & 0.030 \\
 & Reddit & 0.054 & 47.939 & 0.482 & 0.030 \\
 & Twitter & 0.052 & 45.911 & 0.501 & 0.036 \\
 & YouTube & 0.051 & 43.599 & 0.522 & 0.033 \\
\bottomrule
\end{tabular}
\end{table}

\section*{Conclusion}

The spread of information on social media platforms and how the public receives it often aligns with the principles of agenda-setting theory, showcasing what issues gain prominence and trigger discussions in digital public realms. This study, by comparing the discourse around ChatGPT with the discussions on COVID-19 vaccines, highlights the notable differences in information diffusion depending on the nature and global relevance of the topics.
In more detail, the analysis of online discussion dynamics regarding Large Language Models, particularly ChatGPT, across different social media platforms and news articles reveals a complex picture. The posts that gain the most interaction usually align with users' majority interest on each platform, possibly reflecting the selective exposure theory. Meanwhile, the prominence of ChatGPT discussions following its release showcases the agenda-setting potential of new technological developments. Similarly, the rapid and extensive spread of discussions around COVID-19 vaccines highlights the urgency and global concern tied to the pandemic, aligning with traditional agenda-setting models where pressing issues dominate public discourse. 
Our cross-platform analysis reveals distinct engagement dynamics across platforms, emphasizing the importance of the platform environment and user base in digital agenda-setting. 
Moreover, the difference in information spread between these two topics underlines the unique opportunity that massive global events offer in studying societal engagement on digital platforms. With their real-time and global reach, social media have reshaped collective participation in response to global events and opened new ways to analyze these engagements through data.
This study highlights a crucial aspect of our digital age: the complex interplay between the nature of information, the dynamics of social media platforms, and the collective engagement of the user base in agenda-setting, particularly during globally significant events. Furthermore, the insights from such analyses present a vital pathway to explore how social media platforms can be leveraged to promote informed discourse and engagement.
As the frontier of AI technologies like ChatGPT continues to advance, comprehending the discourse surrounding them, how it is shaped, and how it disseminates across various platforms becomes crucial. 

Despite the fears surrounding the misuse of such models, especially in the potential creation of misinformation, the discussions generally centered around the technical aspects and potential usage of LLMs, such as in creative writing – an important insight for AI developers and policymakers. Future research could investigate such patterns in other areas of AI or technology innovations to provide a broader understanding of discourse dynamics around emerging technologies. Furthermore, other research efforts can cover more platforms, analyzing different temporal frames and comparing other significant global events. This will enhance our understanding of digital agenda-setting and inform effective communication, public engagement, and policy-making strategies in our increasingly interconnected digital society.
These insights can guide AI and technology companies, educators, and policymakers in effective communication and anticipating public responses to new technological developments. The lessons learned here underscore the importance of a varied communication strategy tailored to the dynamics of different platforms. This study offers a crucial lens into the communication dynamics of controversial or complex digital developments, providing a vital foundation for future research and practice.

\section*{Methods}
\subsection*{Data Collection}
Our approach to data collection involved utilizing two specific keywords: "OpenAI" and "ChatGPT", in a case-insensitive manner, over a time span from November 25, 2022, to February 25, 2023. When it came to the information related to the COVID-19 vaccine, we built upon the previously accumulated data\cite{valensise2021lack}, further expanding this data by using the same keywords reported to extract relevant news articles from the GDELT dataset, as well as posts from Instagram, Reddit and YouTube. Each platform required a unique method to extract the necessary content. For Facebook and Instagram, we employed CrowdTangle\cite{crowdtangle2023}, a tool that offers social media analytics by tracking public content on different social media platforms. In the case of the Reddit dataset, we initially integrated the use of both CrowdTangle and Pushshift. However, we resort to Reddit dump after encountering issues with the Pushshift API.
The Twitter dataset was compiled using its official API, prior to enforcement of rate alterations. As for the YouTube data, this was collected via the YouTube data API, and news articles were retrieved from the GDELT's Global Knowledge Graph (GKG) table by using Google's BigQuery service\footnote{\url{https://blog.gdeltproject.org/gdelt-2-0-our-global-world-in-realtime/}}. Table \ref{data_breakdown} shows a detailed breakdown of the data. Our data collection was not without limitations. For instance, CrowdTangle can only track data to a certain limit \footnote{\url{https://help.crowdtangle.com/en/articles/1140930-what-data-is-crowdtangle-tracking}}. As for the GDELT news collection, it was dependent on whether the keyword was present in the article's URL. 

\begin{table}[ht!]
\centering
\caption{Data Breakdown}
\label{data_breakdown}
\begin{tabular}{c c c c c}
\toprule
\textbf{Topic} & \textbf{Platform} & \textbf{Posts} & \textbf{Users} & \textbf{Period} \\
\midrule
\multirow{6}{*}{\centering ChatGPT} & Facebook & 220,750 & 72,011 & \multirow{6}{*}{\centering 11/25/2022 - 02/25/2023} \\
 & Instagram & 19,119 & 9,584 & \\
 & News & 17,773 & - & \\
 & Reddit & 25,278 & 15,403 & \\
 & Twitter & 2,597,347 & 982,111 & \\
 & Youtube & 26,676 & 11,433 & \\
\midrule
\multirow{5}{*}{\centering Vaccine} & Facebook & 5,055,483 & 596,705 & \multirow{5}{*}{\centering 11/01/2020 - 02/01/2021} \\
 & Instagram & 434,824 & 152,136 & \\
 & News & 551,832 & - & \\
 & Reddit & 409,692 & 137,099 & \\
 & Twitter & 3,146,019 & 1,104,221 & \\
 & Youtube & 66,702 & 12,996 & \\
\bottomrule
\end{tabular}
\end{table}

There is no doubt that all datasets contain some amount of spam. However, in the case of Facebook, the presence of spam was apparent from the begining of our analysis. Consequently, we decided to filter out these spam posts that were hijacking ChatGPT hashtag, which in turn reduced the size of Facebook's dataset to one-fourth (800k to 200k). This process was based on the simultaneous usage of these hashtags: \#reeel, \#cr7, \#chatgpt, \#fyp, \#viral (see SI fo further details).

\subsection*{Topic Descriptions}

In this section, we outline the key topics identified from our analysis and provide a description for each, highlighting the main discussions shared by users. These topics are as follows:
\begin{itemize}
    \item The \textit{AI Growth} topic covers comments highlighting the rapid increase in users drawn to OpenAI's ChatGPT technology.  
    \item \textit{AI Rivalry} captures comments on the competitive positions taken by major tech entities like Google, Baidu, Microsoft, Meta, Amazon, and Nvidia towards the rise of OpenAI's ChatGPT.
    \item Topic \textit{Access ChatGPT} is about discussions surrounding the methods of accessing ChatGPT service. It consists of conversations about country-specific restrictions, the use of alternative means like fake phone numbers, and details related to Plus subscribers.
    \item  The fourth topic, \textit{Creative Writing}, spans a broad spectrum of artistic expression. It encapsulates discussions and requests related to various forms of written art, such as poetry, songs, screenplays, in addition to crafting jokes, designing itineraries, and writing books. 
    \item The topic of \textit{Cryptocurrency} focuses on discussions about digital currencies, mainly Bitcoin, Ethereum, and Dogecoin, and their price predictions. While this topic could have been combined with "financial discussions", it was kept separate due to its significant size. Thematically, cryptocurrency is also separate from the realm of finance. 
    \item \textit{Education} is one of the main topics discussed in our datasets and covers a wide range of sub-topics. Discussions often touch on issues like plagiarism and AI-generated essays, students leveraging LLMs to cheat on assignments, and schools taking measures to ban the use of such models. There's also a keen interest in how LLMs perform on exams and their utility in answering questions in fields like math and physics. The topic further extends to language learning and LLMs' role in translation. Notably, this topic has been the focus of other research\cite{ChatGPT_for_good,gptinhighered}.
    \item  The \textit{Entertainment} topic branches into three main sub-categories: recipe ideas, sports, and gaming. The first category captures requests related to new food or cocktail ideas. In sports, users often seek ChatGPT's insights on players, teams, and strategies across various leagues like the NBA, F1, and football. On the gaming front, discussions revolve around enhancing game designs and tips for achieving higher scores in specific games.
    \item A significant portion of the comments are dedicated to \textit{Financial Discussions} topic, emphasizing the convergence of AI and finance. Users mainly discussed AI's influence in marketing, stock trading, and catalyzing business growth in addition to the potential of LLMs to revolutionize entrepreneurship and enhance SEO practices. Discussions often touched on optimizing business strategies and leveraging ChatGPT to boost sales. The potential impact of integrating LLMs in the financial sector has been discussed by many studies\cite{finance2023chatgpt}.
    \item The topic \textit{Health} captures comments about well-being and medical matters. Users discussed various medical issues; therapeutic conversations with the bot, personality assessments, exercise routines, and relationship insights. The significance of health discussions in the context of AI has been detailed in other studies\cite{albahri2023systematic, kumar2023artificial}. 
    \item \textit{Image Generation} centers on comments about the use of advanced tools for creating visuals. Users discussed tools like Midjourney, Stable Diffusion, and DALLE-2, highlighting their capabilities and applications in crafting compelling images.
    \item The \textit{Job Market} topic captures comments about the impact of LLMs on the job market. Users discussed both the positive aspects, such as enhanced productivity and innovation, and concerns regarding the potential of human job displacement\cite{moll2022uneven,whitecollarjobs}. The dual-edged role of automation in recruitment was also highlighted, with candidates using it to refine applications and employers leveraging it for assessment. 
    \item Topic \textit{Public Figures} included comments centered on leading figures in tech and business. Notable figures discussed include Sam Altman, Elon Musk, Stephen Wolfram, Larry Page, Sergey Brin, Jordan Peterson, Yann LeCun, Lex Fridman, Marc Andreessen, and Bill Gates. 
    \item The \textit{Risks} topic is segmented into two main areas: security concerns and accuracy \& bias. Security concerns cover issues like data privacy, which pertains to the potential misuse of private user data; malicious activities, which involve the exploitation of the model for harmful purposes; and jailbreaking ChatGPT, which refers to unauthorized uses of the model. In accuracy \& bias, posts address societal concerns like misinformation and inherent biases in the model, leading to issues such as culture wars, gender, political and religious biases. Additionally, it touches on broader societal impacts when users discuss contentious subjects such as veganism, climate change, and geopolitical tensions like the Russia-Ukraine conflict.
    \item As the final topic, \textit{Technical Discussions} covers the technical sides of AI and LLMs. Users shared insights on AI tutorials, discussed LLM training techniques, looked into open-source models, and explored different ways that AI can be integrated into various tasks such as bot development and speech processing.
\end{itemize}

\subsection*{Topic Modeling}

In our study, we employed BERTopic~\cite{grootendorst2022bertopic} for our topic modeling tasks which is a technique that combines the capabilities of transformer models, such as BERT (Bidirectional Encoder Representations from Transformers), and traditional topic modeling techniques like Latent Dirichlet Allocation (LDA). The advantage of BERTopic is that it leverages the context-capturing capabilities of transformer models, which are superior to traditional techniques when it comes to understanding semantic meanings of words\cite{whybertopicisbest}. 

Initially, the dataset was filtered to include only English-language comments. For language detection, we employed the xlm-roberta-base-language-detection model\cite{xlmroberta2019unsupervised} for all platforms, with the exception of Twitter, as the raw Twitter data was already sorted by language. This model achieved an average accuracy of 99.6\% on a benchmark of 20 languages\cite{huggingfaceXlmroberta}. Next, we preprocessed the data by removing URLs and stop words to reduce noise. For embedding the sentences, we used the all-MiniLM-L6-v2 model from the Sentence Transformers library\cite{reimers-2020-multilingual-sentence-bert}. The BERTopic parameters were then fine-tuned depending on the size of each social media platform's dataset. 

After applying the BERTopic model to our datasets, we obtained a diverse range of topics. Our aim was to establish a consistent set of topics across all datasets, necessitating a careful review and reorganization of the results. Hence, we post-processed the BERTopic results by performing explanatory mixed methods analysis\cite{fetters2019mixed}. This entailed a qualitative aggregation of the numerous topics identified by BERTopic into a smaller number of general topics (i.e. themes). This process drew from an iterative qualitative refinement approach inspired by grounded theory principles\cite{bergman2021textual}. Given the unsupervised nature of the BERTopic model, it was vital to conduct a thorough evaluation of the interpretability of the generated topics. We engaged in reviewing the keywords provided for each topic (i.e. Interpretability evaluation). Due to the often nuanced nature of these keywords, it was also necessary for some instances to examine a sample of 30 to 50 comments assigned to the topic to gain a deeper understanding of its context and relevance (i.e. Content analysis). To come up with a consistent list of topics across five datasets, we embarked on the iterative process of aligning the topics from each dataset to a standardized set of around 14 common topics. This process was dynamic, as our understanding of the overall topic space deepened with the review of each dataset (i.e. Iterative refinement). This occasionally led to the modification of our set of common topics to encapsulate better the spectrum of themes presented in the data. A short description of these finalized common topics can be found in the topic description section, highlighting the key themes and considerations for each one.

Throughout the process, in order to ensure consistency, we needed to make some compromises which was an inevitable consequence of reducing a high-dimensional space to a lower dimension. We adopted a pragmatic approach to streamline the diverse range of themes that emerged from our data, all revolving around the central theme of artificial intelligence, machine learning, and chat bots. Topics that were too broad and failed to deliver meaningful insight were discarded. We manually assigned clear labels to distinguishable topics; for example, topics characterized by keywords such as "poem - write poem - poetry - write - ask write", "music - song - band - sound - album", "valentines - day - love - valentine - valentines day", "email - cold - cold email - lead - write" were categorized under \textit{Creative Writing}. We labeled topics related to financial discussions, identified by keywords like "marketing - business - customer - product - brand", "stock - ai stock - investor - investorideas - stock directory", "money - make money - make - fiverr - money online", as \textit{Financial Discussions}. However, certain topics proved challenging to categorize due to their inherent complexity or vagueness. For instance, topics characterized by keywords like "artificial - artificial intelligence - intelligence - won't believe - dangerous Kansas", "chatbot - ai chatbot - chatbots - ai - ai chatbots", "language - model - language model - large language - transformer", "chat gpt - gpt - chat - gpt chat - ai chat", "know - need know - ask - question - need" were less straightforward. These topics, often driven by short or overly complex comments, were labeled as outliers and removed from our analysis. We were skeptical of such broad topics, to be more specific, a topic such as Q\&A often revolves around specific subjects, and if the topic of the question and answer isn't detected, then these comments require further processing, especially if they contain screenshots of the conversations, making them outliers in our analysis. We recognize that this approach entails a certain degree of subjectivity and could potentially eliminate some relevant information. 

Despite the superior performance of BERTopic over many conventional topic modeling techniques and the process of manual review and adjustment we used, we must acknowledge the limitations of the process. These limitations arise from the inherent approximations made by the model and the unavoidable subjectivity in human judgment during the labeling process. Our endeavor involved the modeling and manual review of around 1.8 million comments. Regardless of these challenges, we strived to provide a coherent set of topics that offer meaningful insights into our data.

\subsection*{Sentiment Tones for Topics}
In our study to understand the sentiment tone associated with topics across different social media platforms, we combined results from the topic modeling section with analysis from the GDELT dataset. The GDELT Project is a global database that tracks news and provides sentiment data on these articles among other features. We focused on news articles published across the same timeframe. We then selected posts across all platforms that contained a URL matching an entry in the GDELT dataset. GDELT's sentiment analysis is based on two main metrics: the Positive Score --- representing the percentage of words in an article with a positive emotional connotation, which ranges from 0 to +100 --- and the Negative Score --- indicating the percentage of words with a negative connotation, also ranging from 0 to +100. The overall sentiment tone is calculated by subtracting the Negative Score from the Positive Score, producing a range from -100 (very negative) to +100 (very positive), with 0 being neutral. \footnote{\url{http://data.gdeltproject.org/documentation/GDELT-Global_Knowledge_Graph_Codebook-V2.1.pdf}} After merging the topic modeling dataset with GDELT, we identified 31,208 URLs. Each of these URLs is linked to a post or comment referencing a specific news article with an associated topic. Table~\ref{tab:tone_stats} displays the number of URLs for each topic and provides statistics for the sentiment tone distribution, including minimum, maximum, 25\textsuperscript{th} percentile, 75\textsuperscript{th} percentile, and mean values.

\subsection*{Logistic Function}
We apply the logistic function (commonly known as s-curve) to model the growth trajectory of users engaged with the ChatGPT and COVID-19 discussions. In this model that was originally devised to model the population growth \cite{cramer2002origins}, the initial stage of growth is approximately exponential; then, as saturation begins, the growth slows to linear, and at maturity, growth stops. The role of logistic function has been emphasized in new product adaptation\cite{rogers1976new}, transport infrastructures' evolution\cite{grubler1990rise}, and interplay between technological revolutions and financial capital\cite{perez2003technological}. Recently, this function has been integrated into online social network analysis \cite{spann2022applying}, and  adapted to analyze the user's engagement dynamics across different topics \cite{etta2023characterizing}. 
To fit the data, we used a logistic function with the following formula:

\begin{equation}
f_{\alpha, \beta}(t) = \frac{1}{1 + e^{-\alpha(t-\beta)}}
\label{eq:sigmoid}
\end{equation}
where $\alpha$ is the slope and corresponds to the user growth rate, while $\beta$ is the point at which the function attains a value of 0.5, indicating when half of the overall unique users have engaged with the subject. The value of $\alpha$ quantifies how fast the number of users is growing. A higher $\alpha$ means that the user numbers are growing at a faster rate, while lower values indicate a less pronounced growth. The value of $\beta$  measures how long it takes for half of the total unique users to engage. A lower $\beta$ means that it takes less time to reach half of the total unique users, implying quicker engagement. Finally, we utilized the speed index function~\cite{etta2023characterizing}, which measures the normalized area under the curve and is defined as follows:

\begin{equation}
SI(f_{\alpha, \beta}) = \frac{\int_0^T f_{\alpha, \beta}(t)\,dt}{T}
\label{eq:speed}
\end{equation}

This index captures how fast the function arrives at its peak, spanning from 0 to 1. It can be used to compare how quickly user engagement dynamics stabilize across different platforms and topics. If the SI value is high, it means the discussion topic saturates quickly among users. In other words, the topic reaches its peak popularity rapidly and then doesn't grow much after that.
Conversely, a low SI indicates a slower yet constant growth, when users keep joining the conversation for a longer time.

\begin{table}[ht!]
\centering
\caption{Statistical metrics of sentiment tone distribution for each topics}
\begin{tabular}{lrrrrrrrrr}
\toprule
\textbf{General\_topic} & \textbf{Posts} & \textbf{Mean} & \( \sigma \) & \( Q1 \) & \( Q3 \) & \textbf{Min} & \textbf{Max} \\
\midrule
AI Growth & 1743 & 0.26 & 1.60 & -0.61 & 1.44 & -5.48 & 4.70 \\
AI Rivalry & 6887 & -0.16 & 1.81 & -1.26 & 1.00 & -8.82 & 8.93 \\
Access ChatGPT & 731 & 0.06 & 1.42 & -0.66 & 0.60 & -6.13 & 8.93 \\
Creative Writing & 1711 & -0.31 & 1.90 & -1.44 & 0.78 & -8.33 & 7.16 \\
Cryptocurrency & 201 & 0.04 & 1.80 & -1.37 & 1.46 & -5.97 & 5.45 \\
Education & 5842 & -0.70 & 1.68 & -1.55 & 0.29 & -8.00 & 7.16 \\
Entertainment & 131 & 0.03 & 1.47 & 0.50 & 0.49 & -3.46 & 5.73 \\
Financial Discussions & 1522 & 0.69 & 1.51 & 0.08 & 1.44 & -8.61 & 6.56 \\
Health & 757 & 0.16 & 1.58 & -0.49 & 0.89 & -6.67 & 5.90 \\
Image Generation & 491 & 1.58 & 1.88 & 0.08 & 3.83 & -3.99 & 6.34 \\
Job Market & 2083 & 0.08 & 1.92 & -0.95 & 1.43 & -8.33 & 8.00 \\
Public Figures & 449 & 0.60 & 1.89 & -0.13 & 1.77 & -6.10 & 11.00 \\
Risks & 5177 & -1.21 & 1.93 & -2.59 & 0.00 & -20.69 & 8.71 \\
Technical Discussions & 3483 & 0.45 & 2.16 & -0.52 & 1.45 & -8.33 & 7.55 \\
\bottomrule
\end{tabular}
\label{tab:tone_stats}
\end{table}

\vspace{1cm}

\section*{Data and Code Availability}
The code repository for this paper can be found at \url{https://github.com/shayanalipour/chatgpt_vs_vaccine}. We are unable to share the raw data obtained from CrowdTangle\footnote{\url{https://help.crowdtangle.com/en/articles/4558716-understanding-and-citing-crowdtangle-data}} but any researcher can gain access to CrowdTangle platform upon request. 
Post IDs are available for Twitter, YouTube, and Reddit data, as specified by the platforms' guidelines.
The raw data from GDELT for both topics is available in addition to the aggregated data for the number of daily posts, interactions, and unique user count for each topic platform. 
We also provide the topic modeling information and news articles associated with each topic. These data will be available on OSF repository upon acceptance of the paper.


\begin{thebibliography}{67}
\providecommand{\natexlab}[1]{#1}
\providecommand{\url}[1]{\texttt{#1}}
\expandafter\ifx\csname urlstyle\endcsname\relax
  \providecommand{\doi}[1]{doi: #1}\else
  \providecommand{\doi}{doi: \begingroup \urlstyle{rm}\Url}\fi

\bibitem[Albahri et~al.(2023)Albahri, Duhaim, Fadhel, Alnoor, Baqer, Alzubaidi, Albahri, Alamoodi, Bai, Salhi, et~al.]{albahri2023systematic}
AS~Albahri, Ali~M Duhaim, Mohammed~A Fadhel, Alhamzah Alnoor, Noor~S Baqer, Laith Alzubaidi, OS~Albahri, AH~Alamoodi, Jinshuai Bai, Asma Salhi, et~al.
\newblock A systematic review of trustworthy and explainable artificial intelligence in healthcare: Assessment of quality, bias risk, and data fusion.
\newblock \emph{Information Fusion}, 2023.

\bibitem[Bail et~al.(2018)Bail, Argyle, Brown, Bumpus, Chen, Hunzaker, Lee, Mann, Merhout, and Volfovsky]{bail2018exposure}
Christopher~A Bail, Lisa~P Argyle, Taylor~W Brown, John~P Bumpus, Haohan Chen, MB~Fallin Hunzaker, Jaemin Lee, Marcus Mann, Friedolin Merhout, and Alexander Volfovsky.
\newblock Exposure to opposing views on social media can increase political polarization.
\newblock \emph{Proceedings of the National Academy of Sciences}, 115\penalty0 (37):\penalty0 9216--9221, 2018.

\bibitem[Bergman(2021)]{bergman2021textual}
Manfred~Max Bergman.
\newblock Textual and audiovisual analyses within a mixed methods framework.
\newblock \emph{SAGE handbook of mixed methods in social \& behavioral research}, page 379, 2021.

\bibitem[Bessi et~al.(2015)Bessi, Coletto, Davidescu, Scala, Caldarelli, and Quattrociocchi]{bessi2015science}
Alessandro Bessi, Mauro Coletto, George~Alexandru Davidescu, Antonio Scala, Guido Caldarelli, and Walter Quattrociocchi.
\newblock Science vs conspiracy: Collective narratives in the age of misinformation.
\newblock \emph{PloS one}, 10\penalty0 (2):\penalty0 e0118093, 2015.

\bibitem[Bey(2023)]{whitecollarjobs}
Matthew Bey.
\newblock The rise of chatgpt and the future of white-collar jobs.
\newblock \emph{On Geopolitics}, page~3, 2023.
\newblock URL \url{https://search-ebscohost-com.ezproxy.aur.edu/login.aspx?direct=true&db=bth&AN=163315559&site=ehost-live}.

\bibitem[Briand et~al.(2021)Briand, Cinelli, Nguyen, Lewis, Prybylski, Valensise, Colizza, Tozzi, Perra, Baronchelli, et~al.]{briand2021infodemics}
Sylvie~C Briand, Matteo Cinelli, Tim Nguyen, Rosamund Lewis, Dimitri Prybylski, Carlo~M Valensise, Vittoria Colizza, Alberto~Eugenio Tozzi, Nicola Perra, Andrea Baronchelli, et~al.
\newblock Infodemics: A new challenge for public health.
\newblock \emph{Cell}, 184\penalty0 (25):\penalty0 6010--6014, 2021.

\bibitem[Brown et~al.(2020)Brown, Mann, Ryder, Subbiah, Kaplan, Dhariwal, Neelakantan, Shyam, Sastry, Askell, Agarwal, Herbert-Voss, Krueger, Henighan, Child, Ramesh, Ziegler, Wu, Winter, Hesse, Chen, Sigler, Litwin, Gray, Chess, Clark, Berner, McCandlish, Radford, Sutskever, and Amodei]{llmIntro}
Tom Brown, Benjamin Mann, Nick Ryder, Melanie Subbiah, Jared~D Kaplan, Prafulla Dhariwal, Arvind Neelakantan, Pranav Shyam, Girish Sastry, Amanda Askell, Sandhini Agarwal, Ariel Herbert-Voss, Gretchen Krueger, Tom Henighan, Rewon Child, Aditya Ramesh, Daniel Ziegler, Jeffrey Wu, Clemens Winter, Chris Hesse, Mark Chen, Eric Sigler, Mateusz Litwin, Scott Gray, Benjamin Chess, Jack Clark, Christopher Berner, Sam McCandlish, Alec Radford, Ilya Sutskever, and Dario Amodei.
\newblock Language models are few-shot learners.
\newblock In H.~Larochelle, M.~Ranzato, R.~Hadsell, M.F. Balcan, and H.~Lin, editors, \emph{Advances in Neural Information Processing Systems}, volume~33, pages 1877--1901. Curran Associates, Inc., 2020.
\newblock URL \url{https://proceedings.neurips.cc/paper_files/paper/2020/file/1457c0d6bfcb4967418bfb8ac142f64a-Paper.pdf}.

\bibitem[Burgess and Green(2018)]{burgess2018youtube}
Jean Burgess and Joshua Green.
\newblock \emph{YouTube: Online video and participatory culture}.
\newblock John Wiley \& Sons, 2018.

\bibitem[Chong and Druckman(2007)]{chong2007framing}
Dennis Chong and James~N Druckman.
\newblock Framing theory.
\newblock \emph{Annu. Rev. Polit. Sci.}, 10:\penalty0 103--126, 2007.

\bibitem[Cinelli et~al.(2020{\natexlab{a}})Cinelli, Brugnoli, Schmidt, Zollo, Quattrociocchi, and Scala]{cinelli2020selective}
Matteo Cinelli, Emanuele Brugnoli, Ana~Lucia Schmidt, Fabiana Zollo, Walter Quattrociocchi, and Antonio Scala.
\newblock Selective exposure shapes the facebook news diet.
\newblock \emph{PloS one}, 15\penalty0 (3):\penalty0 e0229129, 2020{\natexlab{a}}.

\bibitem[Cinelli et~al.(2020{\natexlab{b}})Cinelli, Quattrociocchi, Galeazzi, Valensise, Brugnoli, Schmidt, Zola, Zollo, and Scala]{cinelli2020covid}
Matteo Cinelli, Walter Quattrociocchi, Alessandro Galeazzi, Carlo~Michele Valensise, Emanuele Brugnoli, Ana~Lucia Schmidt, Paola Zola, Fabiana Zollo, and Antonio Scala.
\newblock The covid-19 social media infodemic.
\newblock \emph{Scientific reports}, 10\penalty0 (1):\penalty0 1--10, 2020{\natexlab{b}}.

\bibitem[Cinelli et~al.(2021)Cinelli, De~Francisci~Morales, Galeazzi, Quattrociocchi, and Starnini]{cinelli2021echo}
Matteo Cinelli, Gianmarco De~Francisci~Morales, Alessandro Galeazzi, Walter Quattrociocchi, and Michele Starnini.
\newblock The echo chamber effect on social media.
\newblock \emph{Proceedings of the National Academy of Sciences}, 118\penalty0 (9):\penalty0 e2023301118, 2021.

\bibitem[Conneau et~al.(2019)Conneau, Khandelwal, Goyal, Chaudhary, Wenzek, Guzm{\'a}n, Grave, Ott, Zettlemoyer, and Stoyanov]{xlmroberta2019unsupervised}
Alexis Conneau, Kartikay Khandelwal, Naman Goyal, Vishrav Chaudhary, Guillaume Wenzek, Francisco Guzm{\'a}n, Edouard Grave, Myle Ott, Luke Zettlemoyer, and Veselin Stoyanov.
\newblock Unsupervised cross-lingual representation learning at scale.
\newblock \emph{arXiv preprint arXiv:1911.02116}, 2019.

\bibitem[Cramer(2002)]{cramer2002origins}
Jan~Salomon Cramer.
\newblock The origins of logistic regression.
\newblock \emph{Tinbergen Institute Working Paper}, 2002.

\bibitem[De~Angelis et~al.(2023)De~Angelis, Baglivo, Arzilli, Privitera, Ferragina, Tozzi, and Rizzo]{infodemic2023chatgpt}
Luigi De~Angelis, Francesco Baglivo, Guglielmo Arzilli, Gaetano~Pierpaolo Privitera, Paolo Ferragina, Alberto~Eugenio Tozzi, and Caterina Rizzo.
\newblock Chatgpt and the rise of large language models: the new ai-driven infodemic threat in public health.
\newblock \emph{Frontiers in Public Health}, 11:\penalty0 1166120, 2023.

\bibitem[Del~Vicario et~al.(2016)Del~Vicario, Bessi, Zollo, Petroni, Scala, Caldarelli, Stanley, and Quattrociocchi]{del2016spreading}
Michela Del~Vicario, Alessandro Bessi, Fabiana Zollo, Fabio Petroni, Antonio Scala, Guido Caldarelli, H~Eugene Stanley, and Walter Quattrociocchi.
\newblock The spreading of misinformation online.
\newblock \emph{Proceedings of the national academy of Sciences}, 113\penalty0 (3):\penalty0 554--559, 2016.

\bibitem[Derner and Batisti{\v{c}}(2023)]{chatgpt_security_risk}
Erik Derner and Kristina Batisti{\v{c}}.
\newblock Beyond the safeguards: Exploring the security risks of chatgpt.
\newblock \emph{arXiv preprint arXiv:2305.08005}, 2023.

\bibitem[Duarte(2023)]{explodingtopics}
Fabio Duarte.
\newblock Number of chatgpt users, 2023.
\newblock URL \url{https://explodingtopics.com/blog/chatgpt-users#growth}.

\bibitem[Dziri et~al.(2023)Dziri, Lu, Sclar, Li, Jian, Lin, West, Bhagavatula, Bras, Hwang, et~al.]{dziri2023faith}
Nouha Dziri, Ximing Lu, Melanie Sclar, Xiang~Lorraine Li, Liwei Jian, Bill~Yuchen Lin, Peter West, Chandra Bhagavatula, Ronan~Le Bras, Jena~D Hwang, et~al.
\newblock Faith and fate: Limits of transformers on compositionality.
\newblock \emph{arXiv preprint arXiv:2305.18654}, 2023.

\bibitem[Egger and Yu(2022)]{whybertopicisbest}
Roman Egger and Joanne Yu.
\newblock A topic modeling comparison between lda, nmf, top2vec, and bertopic to demystify twitter posts.
\newblock \emph{Frontiers in sociology}, 7:\penalty0 886498, 2022.

\bibitem[Entman(1993)]{entman1993framing}
Robert~M Entman.
\newblock Framing: Toward clarification of a fractured paradigm.
\newblock \emph{Journal of communication}, 43\penalty0 (4):\penalty0 51--58, 1993.

\bibitem[Etta et~al.(2023)Etta, Sangiorgio, Di~Marco, Avalle, Scala, Cinelli, and Quattrociocchi]{etta2023characterizing}
Gabriele Etta, Emanuele Sangiorgio, Niccol{\`o} Di~Marco, Michele Avalle, Antonio Scala, Matteo Cinelli, and Walter Quattrociocchi.
\newblock Characterizing engagement dynamics across topics on facebook.
\newblock \emph{Plos one}, 18\penalty0 (6):\penalty0 e0286150, 2023.

\bibitem[Falkenberg et~al.(2022)Falkenberg, Galeazzi, Torricelli, Di~Marco, Larosa, Sas, Mekacher, Pearce, Zollo, Quattrociocchi, et~al.]{falkenberg2022growing}
Max Falkenberg, Alessandro Galeazzi, Maddalena Torricelli, Niccol{\`o} Di~Marco, Francesca Larosa, Madalina Sas, Amin Mekacher, Warren Pearce, Fabiana Zollo, Walter Quattrociocchi, et~al.
\newblock Growing polarization around climate change on social media.
\newblock \emph{Nature Climate Change}, 12\penalty0 (12):\penalty0 1114--1121, 2022.

\bibitem[Fetters(2019)]{fetters2019mixed}
Michael~D Fetters.
\newblock \emph{The mixed methods research workbook: Activities for designing, implementing, and publishing projects}, volume~7.
\newblock Sage Publications, 2019.

\bibitem[Gilardi et~al.(2023)Gilardi, Alizadeh, and Kubli]{gilardi2023chatgpt}
Fabrizio Gilardi, Meysam Alizadeh, and Ma{\"e}l Kubli.
\newblock Chatgpt outperforms crowd-workers for text-annotation tasks.
\newblock \emph{arXiv preprint arXiv:2303.15056}, 2023.

\bibitem[Gozalo-Brizuela and Garrido-Merchan(2023)]{aimodelreviews}
Roberto Gozalo-Brizuela and Eduardo~C Garrido-Merchan.
\newblock Chatgpt is not all you need. a state of the art review of large generative ai models.
\newblock \emph{arXiv preprint arXiv:2301.04655}, 2023.

\bibitem[Grootendorst(2022)]{grootendorst2022bertopic}
Maarten Grootendorst.
\newblock Bertopic: Neural topic modeling with a class-based tf-idf procedure.
\newblock \emph{arXiv preprint arXiv:2203.05794}, 2022.

\bibitem[Grubler(1990)]{grubler1990rise}
Arnulf Grubler.
\newblock \emph{The rise and fall of infrastructures: dynamics of evolution and technological change in transport}.
\newblock Physica-Verlag, 1990.

\bibitem[Gupta et~al.(2023)Gupta, Akiri, Aryal, Parker, and Praharaj]{threatGPT}
Maanak Gupta, CharanKumar Akiri, Kshitiz Aryal, Eli Parker, and Lopamudra Praharaj.
\newblock From chatgpt to threatgpt: Impact of generative ai in cybersecurity and privacy.
\newblock \emph{arXiv preprint arXiv:2307.00691}, 2023.

\bibitem[Haque et~al.(2022)Haque, Dharmadasa, Sworna, Rajapakse, and Ahmad]{early2022adapt}
Mubin~Ul Haque, Isuru Dharmadasa, Zarrin~Tasnim Sworna, Roshan~Namal Rajapakse, and Hussain Ahmad.
\newblock "i think this is the most disruptive technology": Exploring sentiments of chatgpt early adopters using twitter data, 2022.

\bibitem[HuggingFace(2021)]{huggingfaceXlmroberta}
HuggingFace.
\newblock xlm-roberta-base-language-detection, 2021.
\newblock URL \url{https://huggingface.co/papluca/xlm-roberta-base-language-detection}.

\bibitem[Javaid et~al.(2023)Javaid, Haleem, and Singh]{javaid2023chatgpt}
Mohd Javaid, Abid Haleem, and Ravi~Pratap Singh.
\newblock Chatgpt for healthcare services: An emerging stage for an innovative perspective.
\newblock \emph{BenchCouncil Transactions on Benchmarks, Standards and Evaluations}, 3\penalty0 (1):\penalty0 100105, 2023.

\bibitem[Ji et~al.(2023)Ji, Lee, Frieske, Yu, Su, Xu, Ishii, Bang, Madotto, and Fung]{hallucination}
Ziwei Ji, Nayeon Lee, Rita Frieske, Tiezheng Yu, Dan Su, Yan Xu, Etsuko Ishii, Ye~Jin Bang, Andrea Madotto, and Pascale Fung.
\newblock Survey of hallucination in natural language generation.
\newblock \emph{ACM Comput. Surv.}, 55\penalty0 (12), mar 2023.
\newblock ISSN 0360-0300.

\bibitem[Kasneci et~al.(2023)Kasneci, Se{\ss}ler, K{\"u}chemann, Bannert, Dementieva, Fischer, Gasser, Groh, G{\"u}nnemann, H{\"u}llermeier, et~al.]{ChatGPT_for_good}
Enkelejda Kasneci, Kathrin Se{\ss}ler, Stefan K{\"u}chemann, Maria Bannert, Daryna Dementieva, Frank Fischer, Urs Gasser, Georg Groh, Stephan G{\"u}nnemann, Eyke H{\"u}llermeier, et~al.
\newblock Chatgpt for good? on opportunities and challenges of large language models for education.
\newblock \emph{Learning and Individual Differences}, 103:\penalty0 102274, 2023.

\bibitem[Koonchanok et~al.(2023)Koonchanok, Pan, and Jang]{koonchanok2023tracking}
Ratanond Koonchanok, Yanling Pan, and Hyeju Jang.
\newblock Tracking public attitudes toward chatgpt on twitter using sentiment analysis and topic modeling, 2023.

\bibitem[Kortemeyer(2023)]{chatgptPandemic}
Gerd Kortemeyer.
\newblock Artificial intelligence is not a pandemic, 2023.
\newblock URL \url{https://ethz.ch/en/news-and-events/eth-news/news/2023/04/artificial-intelligence-is-not-a-pandemic.html}.

\bibitem[Kubin and von Sikorski(2021)]{kubin2021role}
Emily Kubin and Christian von Sikorski.
\newblock The role of (social) media in political polarization: a systematic review.
\newblock \emph{Annals of the International Communication Association}, 45\penalty0 (3):\penalty0 188--206, 2021.

\bibitem[Kumar et~al.(2023)Kumar, Chauhan, and Awasthi]{kumar2023artificial}
Pranjal Kumar, Siddhartha Chauhan, and Lalit~Kumar Awasthi.
\newblock Artificial intelligence in healthcare: review, ethics, trust challenges \& future research directions.
\newblock \emph{Engineering Applications of Artificial Intelligence}, 120:\penalty0 105894, 2023.

\bibitem[Leetaru and Schrodt(2013)]{leetaru2013gdelt}
Kalev Leetaru and Philip~A Schrodt.
\newblock Gdelt: Global data on events, location, and tone, 1979--2012.
\newblock In \emph{ISA annual convention}, number~4, pages 1--49. Citeseer, 2013.

\bibitem[Leetaru(2015)]{leetaru2015mining}
Kalev~Hannes Leetaru.
\newblock Mining libraries: Lessons learned from 20 years of massive computing on the world’s information.
\newblock \emph{Information Services \& Use}, 35\penalty0 (1-2):\penalty0 31--50, 2015.

\bibitem[Leiter et~al.(2023)Leiter, Zhang, Chen, Belouadi, Larionov, Fresen, and Eger]{leiter2023chatgpt}
Christoph Leiter, Ran Zhang, Yanran Chen, Jonas Belouadi, Daniil Larionov, Vivian Fresen, and Steffen Eger.
\newblock Chatgpt: A meta-analysis after 2.5 months, 2023.

\bibitem[Lopez-Lira and Tang(2023)]{finance2023chatgpt}
Alejandro Lopez-Lira and Yuehua Tang.
\newblock Can chatgpt forecast stock price movements? return predictability and large language models, 2023.

\bibitem[Mahowald et~al.(2023)Mahowald, Ivanova, Blank, Kanwisher, Tenenbaum, and Fedorenko]{mahowald2023dissociating}
Kyle Mahowald, Anna~A Ivanova, Idan~A Blank, Nancy Kanwisher, Joshua~B Tenenbaum, and Evelina Fedorenko.
\newblock Dissociating language and thought in large language models: a cognitive perspective.
\newblock \emph{arXiv preprint arXiv:2301.06627}, 2023.

\bibitem[McCombs and Shaw(1972)]{mccombs1972agenda}
Maxwell~E McCombs and Donald~L Shaw.
\newblock The agenda-setting function of mass media.
\newblock \emph{Public opinion quarterly}, 36\penalty0 (2):\penalty0 176--187, 1972.

\bibitem[Messing and Westwood(2014)]{messing2014selective}
Solomon Messing and Sean~J Westwood.
\newblock Selective exposure in the age of social media: Endorsements trump partisan source affiliation when selecting news online.
\newblock \emph{Communication research}, 41\penalty0 (8):\penalty0 1042--1063, 2014.

\bibitem[Moll et~al.(2022)Moll, Rachel, and Restrepo]{moll2022uneven}
Benjamin Moll, Lukasz Rachel, and Pascual Restrepo.
\newblock Uneven growth: automation's impact on income and wealth inequality.
\newblock \emph{Econometrica}, 90\penalty0 (6):\penalty0 2645--2683, 2022.

\bibitem[Perez(2003)]{perez2003technological}
Carlota Perez.
\newblock \emph{Technological revolutions and financial capital}.
\newblock Edward Elgar Publishing, 2003.

\bibitem[Quattrociocchi et~al.(2014)Quattrociocchi, Caldarelli, and Scala]{quattrociocchi2014opinion}
Walter Quattrociocchi, Guido Caldarelli, and Antonio Scala.
\newblock Opinion dynamics on interacting networks: media competition and social influence.
\newblock \emph{Scientific reports}, 4\penalty0 (1):\penalty0 4938, 2014.

\bibitem[Radford et~al.(2019)Radford, Wu, Child, Luan, Amodei, Sutskever, et~al.]{radford2019language}
Alec Radford, Jeffrey Wu, Rewon Child, David Luan, Dario Amodei, Ilya Sutskever, et~al.
\newblock Language models are unsupervised multitask learners.
\newblock \emph{OpenAI blog}, 1\penalty0 (8):\penalty0 9, 2019.

\bibitem[Rasul et~al.(2023)Rasul, Nair, Kalendra, Robin, de~Oliveira~Santini, Ladeira, Sun, Day, Rather, and Heathcote]{gptinhighered}
Tareq Rasul, Sumesh Nair, Diane Kalendra, Mulyadi Robin, Fernando de~Oliveira~Santini, Wagner~Junior Ladeira, Mingwei Sun, Ingrid Day, Raouf~Ahmad Rather, and Liz Heathcote.
\newblock The role of chatgpt in higher education: Benefits, challenges, and future research directions.
\newblock \emph{Journal of Applied Learning and Teaching}, 6\penalty0 (1), 2023.

\bibitem[Ray(2023)]{ray2023chatgpt}
Partha~Pratim Ray.
\newblock Chatgpt: A comprehensive review on background, applications, key challenges, bias, ethics, limitations and future scope.
\newblock \emph{Internet of Things and Cyber-Physical Systems}, 2023.

\bibitem[Reimers and Gurevych(2020)]{reimers-2020-multilingual-sentence-bert}
Nils Reimers and Iryna Gurevych.
\newblock Making monolingual sentence embeddings multilingual using knowledge distillation.
\newblock In \emph{Proceedings of the 2020 Conference on Empirical Methods in Natural Language Processing}. Association for Computational Linguistics, 11 2020.
\newblock URL \url{https://arxiv.org/abs/2004.09813}.

\bibitem[Rogers(1976)]{rogers1976new}
Everett~M Rogers.
\newblock New product adoption and diffusion.
\newblock \emph{Journal of consumer Research}, 2\penalty0 (4):\penalty0 290--301, 1976.

\bibitem[Ruffo et~al.(2023)Ruffo, Semeraro, Giachanou, and Rosso]{newsdiffusion}
Giancarlo Ruffo, Alfonso Semeraro, Anastasia Giachanou, and Paolo Rosso.
\newblock Studying fake news spreading, polarisation dynamics, and manipulation by bots: A tale of networks and language.
\newblock \emph{Computer Science Review}, 47:\penalty0 100531, 2023.
\newblock ISSN 1574-0137.

\bibitem[Scheufele and Tewksbury(2007)]{scheufele2007framing}
Dietram~A Scheufele and David Tewksbury.
\newblock Framing, agenda setting, and priming: The evolution of three media effects models.
\newblock \emph{Journal of communication}, 57\penalty0 (1):\penalty0 9--20, 2007.

\bibitem[Schmidt et~al.(2017)Schmidt, Zollo, Del~Vicario, Bessi, Scala, Caldarelli, Stanley, and Quattrociocchi]{schmidt2017anatomy}
Ana~Luc{\'\i}a Schmidt, Fabiana Zollo, Michela Del~Vicario, Alessandro Bessi, Antonio Scala, Guido Caldarelli, H~Eugene Stanley, and Walter Quattrociocchi.
\newblock Anatomy of news consumption on facebook.
\newblock \emph{Proceedings of the National Academy of Sciences}, 114\penalty0 (12):\penalty0 3035--3039, 2017.

\bibitem[Sears and Freedman(1967)]{sears1967selective}
David~O Sears and Jonathan~L Freedman.
\newblock Selective exposure to information: A critical review.
\newblock \emph{Public Opinion Quarterly}, 31\penalty0 (2):\penalty0 194--213, 1967.

\bibitem[Spann et~al.(2022)Spann, Mead, Maleki, Agarwal, and Williams]{spann2022applying}
Billy Spann, Esther Mead, Maryam Maleki, Nitin Agarwal, and Therese Williams.
\newblock Applying diffusion of innovations theory to social networks to understand the stages of adoption in connective action campaigns.
\newblock \emph{Online Social Networks and Media}, 28:\penalty0 100201, 2022.

\bibitem[Stroud(2008)]{stroud2008media}
Natalie~Jomini Stroud.
\newblock Media use and political predispositions: Revisiting the concept of selective exposure.
\newblock \emph{Political Behavior}, 30:\penalty0 341--366, 2008.

\bibitem[Taecharungroj(2023)]{whatChatGPTdo}
Viriya Taecharungroj.
\newblock What can chatgpt do?; analyzing early reactions to the innovative ai chatbot on twitter.
\newblock \emph{Big Data and Cognitive Computing}, 7\penalty0 (1), 2023.
\newblock ISSN 2504-2289.
\newblock URL \url{https://www.mdpi.com/2504-2289/7/1/35}.

\bibitem[Team(2023)]{crowdtangle2023}
CrowdTangle Team.
\newblock \emph{CrowdTangle}.
\newblock Facebook, Menlo Park, California, United States, 2023.

\bibitem[Turcotte et~al.(2015)Turcotte, York, Irving, Scholl, and Pingree]{socialMediaBetterNews}
Jason Turcotte, Chance York, Jacob Irving, Rosanne~M. Scholl, and Raymond~J. Pingree.
\newblock {News Recommendations from Social Media Opinion Leaders: Effects on Media Trust and Information Seeking}.
\newblock \emph{Journal of Computer-Mediated Communication}, 20\penalty0 (5):\penalty0 520--535, 06 2015.
\newblock ISSN 1083-6101.

\bibitem[Valensise et~al.(2021)Valensise, Cinelli, Nadini, Galeazzi, Peruzzi, Etta, Zollo, Baronchelli, and Quattrociocchi]{valensise2021lack}
Carlo~M. Valensise, Matteo Cinelli, Matthieu Nadini, Alessandro Galeazzi, Antonio Peruzzi, Gabriele Etta, Fabiana Zollo, Andrea Baronchelli, and Walter Quattrociocchi.
\newblock Lack of evidence for correlation between covid-19 infodemic and vaccine acceptance, 2021.

\bibitem[Valensise et~al.(2023)Valensise, Cinelli, and Quattrociocchi]{valensise2023drivers}
Carlo~M Valensise, Matteo Cinelli, and Walter Quattrociocchi.
\newblock The drivers of online polarization: Fitting models to data.
\newblock \emph{Information Sciences}, 642:\penalty0 119152, 2023.

\bibitem[Valmeekam et~al.(2022)Valmeekam, Olmo, Sreedharan, and Kambhampati]{valmeekam2022large}
Karthik Valmeekam, Alberto Olmo, Sarath Sreedharan, and Subbarao Kambhampati.
\newblock Large language models still can't plan (a benchmark for llms on planning and reasoning about change).
\newblock \emph{arXiv preprint arXiv:2206.10498}, 2022.

\bibitem[Zarocostas(2020)]{zarocostas2020fight}
John Zarocostas.
\newblock How to fight an infodemic.
\newblock \emph{The lancet}, 395\penalty0 (10225):\penalty0 676, 2020.

\bibitem[Zollo et~al.(2017)Zollo, Bessi, Del~Vicario, Scala, Caldarelli, Shekhtman, Havlin, and Quattrociocchi]{zollo2017debunking}
Fabiana Zollo, Alessandro Bessi, Michela Del~Vicario, Antonio Scala, Guido Caldarelli, Louis Shekhtman, Shlomo Havlin, and Walter Quattrociocchi.
\newblock Debunking in a world of tribes.
\newblock \emph{PloS one}, 12\penalty0 (7):\penalty0 e0181821, 2017.

\end{thebibliography}

\section*{Acknowledgements}
We would like to thank Abigail Milovancevic, Shahab Mousavi, and Amirhossein Afsharrad for their help and insights in this project.
This work is supported by IRIS Infodemic Coalition (UK government, grant no. SCH-00001-3391), 
SERICS (PE00000014) under the NRRP MUR program funded by the European Union - NextGenerationEU, project CRESP from the Italian Ministry of Health under the program CCM 2022, and PON project “Ricerca e Innovazione” 2014-2020 and project SEED n. SP122184858BEDB3. L.B. acknowledges the support of the Ministry of Science, Technological Development and Innovation of the Republic of Serbia, according to the Agreement on the realization and financing of scientific research.

\section*{Author Contributions Statement}
S.A., A.G. and M.C. collected the data; S.A. analysed the data and produced figures; S.A., A.G., E.S., M.C. and W.Q designed the study; A.G, M.C. and W.Q. supervised the project. All the authors wrote the paper.

\section*{Supporting Information}

\subsection*{COVID-19 Vaccination Dataset}
\begin{figure}[ht!]
    \centering
    \includegraphics[width=0.9\textwidth]{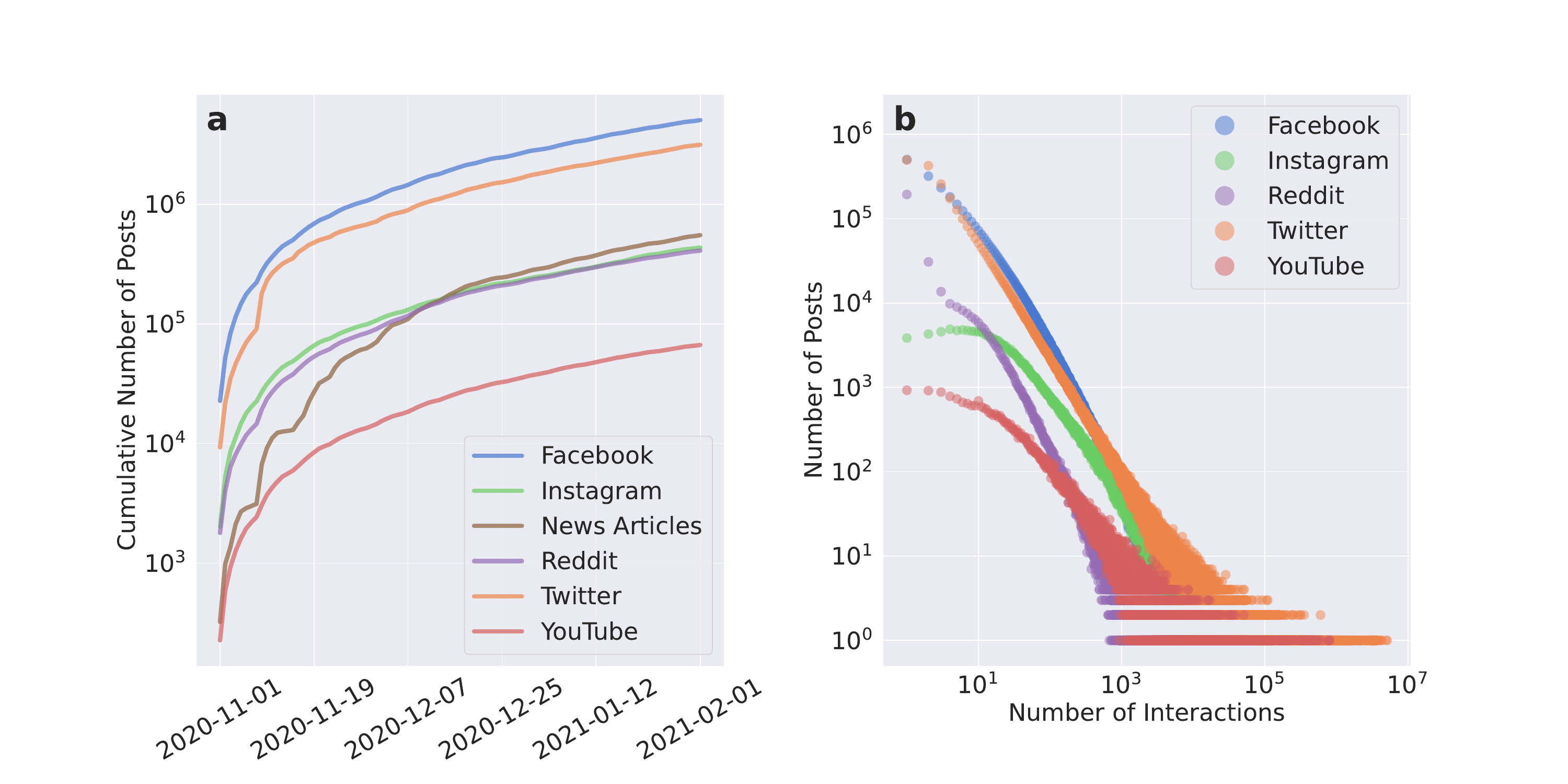}
    \caption{Cumulative number of unique posts about COVID vaccine discussion across various platforms (a) and distribution of interaction volume versus number of posts on different platforms (b).}
    \label{fig:covid_general_info}
\end{figure}

In this section, we present an overview of the characteristics of datasets collected across different platforms related to COVID-19 vaccination debates. Figure \ref{fig:covid_general_info}.a shows the cumulative count of content related to vaccine debates over the span of three months. Figure \ref{fig:covid_general_info}.b presents the distribution of interactions per platform. Interactions are a measure of user engagement with posts on social media platforms, and the definition can vary depending on the platform. On Twitter, interactions are the sum of likes, quotes, retweets, and replies for each post. On Instagram and YouTube, interactions are composed of likes and comments.

\subsection*{Facebook Spam Content}
\begin{figure}[ht!]
    \centering
    \includegraphics[width=0.9\textwidth]{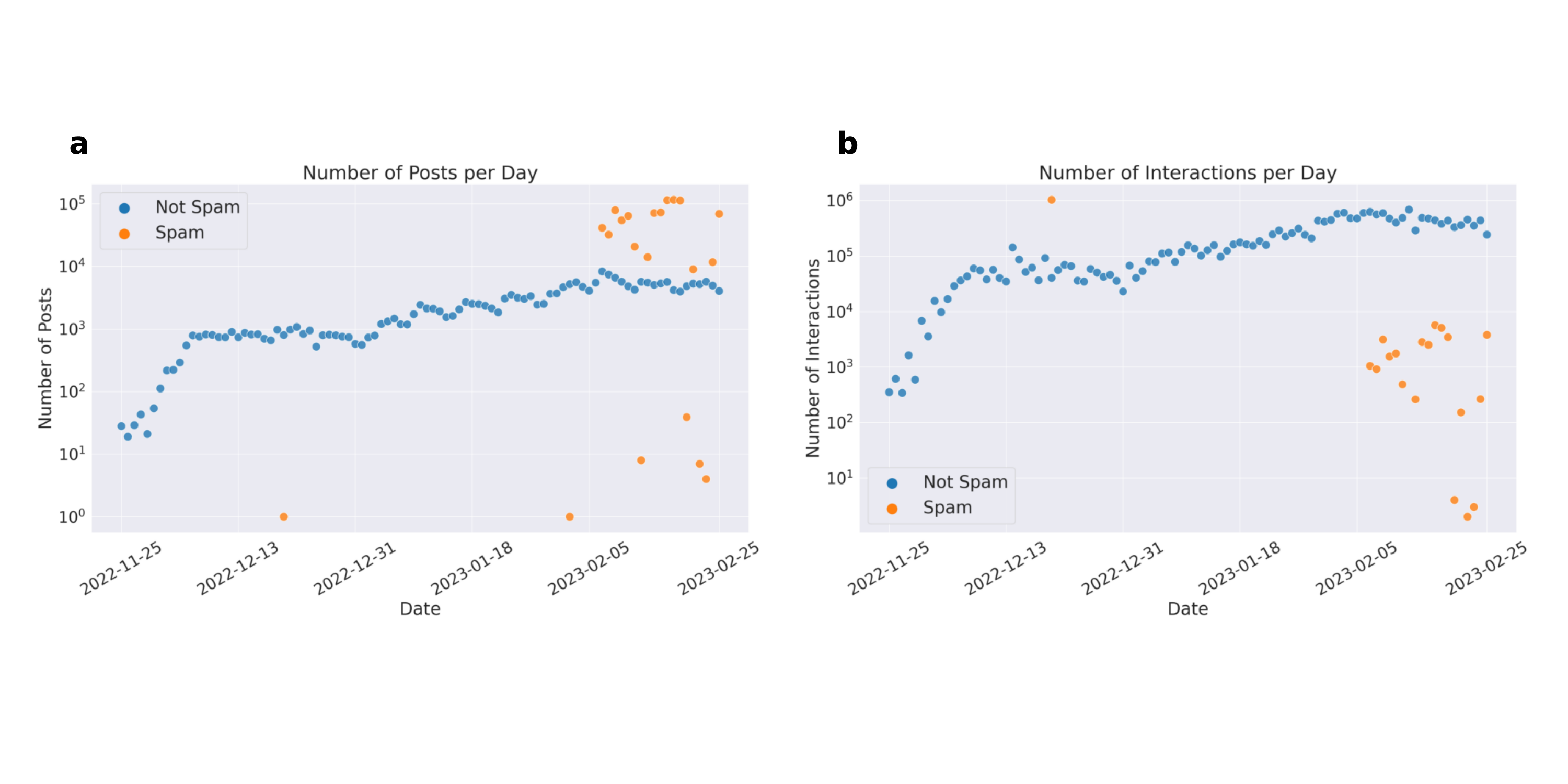}
    \caption{Number of daily posts and interactions for discussions about ChatGPT on Facebook. Content that we flagged as spam shows a totally different behavior.}
    \label{fig:fb_spams}
\end{figure}

Upon analyzing data related to ChatGPT discussions on Facebook, we observed some sharp strange dents in the cumulative number of posts and cumulative number of users. We then checked the dataset and found out there are many posts with similar styles publishing explicit or totally unrelated content in mass numbers that was inevitable to ignore. Our investigation resulted in flagging posts that used either of the following strings in their content: ``\#reels \#chatgpt'', ``Video Funny Amazing \#fyp \#viral'', and ``\#reeel \#cr7 \#chatgpt''.  We then proceed with plotting these flagged posts alongside the normal ones as shown in Figure \ref{fig:fb_spams}. We observed orders of magnitude differences in the number of published posts per day and received interactions for these spam content, indicating that our efforts in removing these spam content were successful.

\end{document}